\catcode`\@=11

\newif\if@fewtab\@fewtabtrue
{\count255=\time\divide\count255 by 60
\xdef\hourmin{\number\count255}
\multiply\count255 by-60\advance\count255 by\time
\xdef\hourmin{\hourmin:\ifnum\count255<10 0\fi\the\count255}}
\def\ps@draft{\let\@mkboth\@gobbletwo
    \def\@oddfoot{\hbox to 7 cm{\tiny \versionno
       \hfil}\hskip -7cm\hfil\rm\thepage \hfil {\tiny\draftdate}}
    \def\@oddhead{}
    \def\@evenhead{}\let\@evenfoot\@oddfoot}
\def\draftdate{\number\month/\number\day/\number\year\ \ \ \hourmin }

\global\def\draftcontrol{0}

\def\draftcite#1{\ifnum\draftcontrol=1#1\else{}\fi}
\def\@lbibitem[#1]#2{\item{}\hskip -3\hbox to 2cm
{\hfil$\scriptstyle\draftcite{#2}$}\hskip
1cm[\@biblabel{#1}]\if@filesw
     {\def\protect##1{\string ##1\space}\immediate
      \write\@auxout{\string\bibcite{#2}{#1}}}\fi\ignorespaces}

\def\@bibitem#1{\item\hskip -3cm \hbox to 2cm
{\hfil {\footnotesize\draftcite{#1}}}\hskip 1cm
\if@filesw \immediate\write\@auxout
       {\string\bibcite{#1}{\the\value{\@listctr}}}\fi\ignorespaces}

\def\citen#1{\if@filesw \immediate\write \@auxout {\string\citation{#1}}\fi%
\@tempcntb\m@ne \let\@h@ld\relax \def\@citea{}%
\@for \@citeb:=#1\do {\@ifundefined {b@\@citeb}%
    {\@h@ld\@citea\@tempcntb\m@ne{\bf ?}%
    \@warning {Citation `\@citeb ' on page \thepage \space undefined}}%
    {\@tempcnta\@tempcntb \advance\@tempcnta\@ne
    \setbox\z@\hbox\bgroup\ifcat0\csname b@\@citeb \endcsname \relax
    \egroup \@tempcntb\number\csname b@\@citeb \endcsname \relax
    \else \egroup \@tempcntb\m@ne \fi \ifnum\@tempcnta=\@tempcntb
    \ifx\@h@ld\relax \edef \@h@ld{\@citea\csname b@\@citeb\endcsname}%
    \else \edef\@h@ld{\hbox{--}\penalty\@highpenalty
    \csname b@\@citeb\endcsname}\fi
    \else \@h@ld\@citea\csname b@\@citeb \endcsname \let\@h@ld\relax \fi}%
\def\@citea{,\penalty\@highpenalty\hskip.13em plus.13em minus.13em}}\@h@ld}
\def\@citex[#1]#2{\@cite{\citen{#2}}{#1}}%
\def\@cite#1#2{\leavevmode\unskip\ifnum\lastpenalty=\z@\penalty\@highpenalty\fi%
  \ [{\multiply\@highpenalty 3 #1%
  \if@tempswa,\penalty\@highpenalty\ #2\fi}]}   %
\makeatother 

\catcode`\@=12

\def\alg           {algebra}
\def\auto          {automorphism}
\def\bc            {boundary condition}
\def\Bc            {Boundary condition}
\def\be            {\begin{equation}}
\def\bearl         {\begin{array}{l}}
\def\bearll        {\begin{array}{ll}}
\def\cft           {conformal field theory}
\def\cfts          {conformal field theories}
\def\Cbf           {{\dl C}}
\def\chii          {{\raisebox{.15em}{$\chi$}}}
\def\class         {classification}

\def\dl            {\mathbb }
\def\dsty          {\displaystyle}
\def\dyd           {Dynkin diagram}
\def\ee            {\end{equation}}
\def\eE            {{\rm e}}
\def\eear          {\end{array}}
\def\eq            {\,{=}\,}
\newcommand\erf[1] {(\ref{#1})}
\newcommand\Erf[2] {(\ref{#1#2})}

\def\furu          {fusion rule}
\def\futnote#1     {\footnote{~#1}\ }
\def\half          {\frac12\,}

\def\ii            {{\rm i}}

\def\iN            {\,{\in}\,}
\long\def\labl#1   {\label{#1}\ee \ifnum\draftcontrol=1
                   \mbox{ }\\[-12 mm]\query{#1}\\[5 mm] \fi}
\long\def\Labl#1#2 {\label{#1#2}\ee\ifnum\draftcontrol=1
                   \mbox{ }\\[-12 mm]\query{#1#2}\\[5 mm] \fi}
\def\mod           {{\ \rm mod\,}}  
\newcommand\mysec[2]{{}\vskip1.3em\noindent{\bf #1.\ #2}\\[.5em]}
\def\Moebius       {M\"obius }
\long\def\query#1{\hskip 0pt{\vadjust{\everypar={}\small\vtop to 0pt{\hbox{}%
     \vskip -13pt\rlap{\hbox to 49.0pc{\hfil{\vtop{\hsize=8pc\tolerance=6000%
     \hfuzz=.5pc\rightskip=0pt plus 3em\noindent#1}}}}\vss}}}}%

\def\stts          {string theories}
\def\syms          {sym\-me\-tries}
\def\tft           {topological field theory}

\def\wzwm          {WZW model}
\def\Zbf           {{\dl Z}}  

\documentclass[12pt]{article} \usepackage{amssymb,amsfonts,latexsym}
\begin{document}


\begin{flushright}  {~} \\[-12mm]
{\sf hep-th/0007174}\\{\sf NIKHEF/2000-019}\\{\sf PAR-LPTHE 00-32}\\
{\sf ETH-TH/00-9}\\{\sf CERN-TH/2000-220}\\[1mm]
{\sf July 2000} \end{flushright}

\begin{center} \vskip 7mm
{\Large\bf BOUNDARIES, CROSSCAPS}\\[3mm]
{\Large\bf AND SIMPLE CURRENTS}\\[16mm]
{\large J.\ Fuchs,$\;^1$ \ L.R.\ Huiszoon,$\;^2$ \ A.N.\ Schellekens,$\;^2$}
\\[2.5mm]
{\large C.\ Schweigert,$\;^3$ \ J.\ Walcher$\;^{4,5}$}
\\[6mm]
$^1\;$ Institutionen f\"or fysik\\
Universitetsgatan 1, \ S\,--\,651\,88\, Karlstad\\[5mm]
$^2\;$ NIKHEF Theory Group\\Postbus 41882, \ NL\,--\,1009~DB\, Amsterdam\\[5mm]
$^3\;$ LPTHE, Universit\'e Paris VI\\
4, place Jussieu, \ F\,--\,75\,252\, Paris\, Cedex 05\\[5mm]
$^4\;$ Institut f\"ur Theoretische Physik \\
ETH H\"onggerberg, \ CH\,--\,8093\, Z\"urich\\[3.5mm]
$^5\;$ Theory Division, CERN\\CH\,--\,1211\, Geneva 23
\end{center}
\vskip 18mm
\begin{quote}{\bf Abstract}\\[1mm]
Universal formulas for the boundary and crosscap coefficients are 
presented, which are valid for all symmetric simple current modifications
of the charge conjugation invariant of any rational conformal field theory.
\end{quote}
\newpage


\mysec1{Boundaries and crosscaps}
In this letter we report progress on the problem of finding boundaries 
and crosscaps for all \cfts\ that can be obtained as simple current 
invariants of a given rational conformal field theory (RCFT). This 
line of research was initiated by Cardy \cite{card9}, who obtained 
the (bulk symmetry-preserving) boundaries in the case where the torus 
partition function is the charge-conjugation invariant. Later, in 
\cite{prss} (see \cite{prad} for a review) the corresponding crosscap 
coefficients were obtained. In a series of subsequent papers 
\cite{prss2,prss3,fuSc5,bppz,fuSc11,huss,bifs,fuSc12,bppz2,huss2,husc},
various more general situations were
considered. In particular one may choose a different Klein bottle
projection and a different modular-invariant partition function (MIPF) 
$\chii_i Z_{ij} \bar\chii_j$ for the bulk theory. 

The basic data one would like to determine are the set $\{m\}$ of Ishibashi 
labels, the set $\{a\}$ of boundary labels, a matrix $B_{m,a}$ of boundary
coefficients, and a vector $\Gamma_m$ of crosscap coefficients.
By ``Ishibashi labels" we mean the labels of the 
Ishibashi states \cite{ishi} that can propagate in the
transverse (closed string) channel. There exists such a label for each 
primary field $i$ that is paired with its conjugate, $i^c$, in the torus
partition function.
A difficulty arises when some of these terms in the
torus partition function have a multiplicity larger than 1. In this
case we must allow for multiplicities in the Ishibashi labels as well;
the degeneracy is precisely given by $Z_{ii^c}$. The Ishibashi labels
then consist of a pair $(i,\alpha)$, where $i$ is a primary field label 
and $\alpha$, which can take $Z_{ii^c}$ different values, takes care
of the degeneracy. 

These data must satisfy a large collection of ``sewing constraints" 
\cite{lewe3,fips,prss3}. Unfortunately, most of them
cannot be checked explicitly because they require detailed
knowledge of fusing matrices, braiding matrices and OPE coefficients. 
However, there exists a set of
simpler constraints, presumably a consequence of the
sewing constraints, but certainly necessary, namely 
the requirement of positivity and integrality of the
partition functions. These partition functions correspond to the
torus, annulus, \Moebius strip and Klein bottle surface.  
Each partition function is a linear combination
of characters $\chii^i$ with arguments that depend on 
the surface under consideration, and with coefficients that depend
on the choice of boundary. These coefficients are given by 
  \be  \bearl 
  A^i_{ab} = \dsty\sum_{j,\alpha,\beta} S^i_j\, B_{(j,\alpha),a}\,
             B_{(j,\beta),b}\,g_j^{\alpha\beta} \,, \\{}\\[-.8em]
  M^i_{a} = \dsty\sum_{j,\alpha,\beta} P^i_j\, B_{(j,\alpha),a}\,
             \Gamma_{(j,\beta)}\,g_j^{\alpha\beta} \,, \\{}\\[-.8em]
  K^i = \dsty\sum_{j,\alpha,\beta} S^i_j\, \Gamma_{(j,\alpha)}\,
             \Gamma_{(j,\beta)}\,g_j^{\alpha\beta}  \eear \ee
for annulus, \Moebius strip and Klein bottle, respectively. 
Here $S$ is the usual modular transformation matrix of the RCFT, and 
$P\eq\sqrt{T}ST^2S\sqrt{T}$ as introduced in \cite{bisa4}.
Further, $g_j$ is a ``metric" in the space of degeneracy labels
of the Ishibashi states belonging to $j$. This is part of the data to be
determined. The torus
partition function is described in terms of a non-negative integer
matrix $Z\eq(Z_{ij})$. All these quantities must be integers, and the 
annulus coefficients as well as the combinations $\half(Z_{ii}+K^i)$ and
$\half (A^i_{aa}+M^i_a)$ (the closed and open string partition
function coefficients) must be non-negative integers. 
Furthermore $A^0_{ab}$, the boundary conjugation matrix (the label ``0"
refers to the vacuum), must be a permutation of order 2. In practice
these conditions have turned out to be very restrictive. In addition
to this one may wish to satisfy the ``completeness conditions" \cite{prss3},
which in the present context is equivalent to requiring that the number
of Ishibashi labels equal the number of boundary labels.

We pause to emphasize that what we wish to obtain is the complete
set of boundaries and crosscaps that possess the symmetry 
$\bar{\mathfrak A}$ of the given unextended theory, even when the 
full bulk symmetry $\mathfrak A$ is larger because of the extension
that is implied by the torus partition function. In other words, even
though we express our results in terms of quantities of the underlying
unextended theory, we
are indeed studying boundaries and crosscaps of a CFT whose chiral
algebra is $\mathfrak A$, not $\bar{\mathfrak A}$. ($\bar{\mathfrak A}$ 
and $\mathfrak A$ coincide if and only if the torus partition function
is a pure \auto\ invariant.) In particular our previous and present 
results include, in this sense, the case of ``symmetry breaking 
boundaries", which preserve only part of the full bulk symmetry.
Also note that in the case of the free boson  
our results amount to finding D-branes (for boundary states) and 
orientifold planes (for crosscaps) that are not space-time filling,
i.e.\ where some directions are Dirichlet, corresponding to the
presence of a non-trivial automorphism for the boundary.

\mysec2{Simple current invariants}
In principle one would like to determine the data listed above for 
arbitrary bulk modular invariants. 
A large subclass of the latter are the simple current invariants. What
we will consider in this paper is in fact the complete class of (symmetric)\,%
 \futnote{In this letter we will demand that the theory exist on unoriented
 surfaces, although it might be possible to relax this condition. This
 requires the torus partition function to be symmetric.} 
simple current modifications of the charge conjugation invariant. 
If the RCFT is real (in the sense that all fields are self-conjugate)
this set nearly exhausts the possibilities, except for a few 
sporadic exceptional invariants. Complex RCFTs possess a second
large set of invariants, namely the simple current modifications of the 
diagonal invariant. The diagonal invariant itself was discussed in 
\cite{bifs} and was found to require additional
data from a suitable orbifold theory. Its simple
current modifications are obviously of interest as well, but they
involve similar complications and are beyond the
scope of this paper.

A complete classification of all simple current invariants
of any RCFT has been achieved in \cite{gasc,krSc}. In various special 
cases, boundaries and crosscaps have already been studied. In 
particular, all cases where
the MIPF is a pure extension of the chiral algebra were dealt
with in \cite{fuSc11,fuSc12} 
as far as the boundaries are concerned. In \cite{husc} the
crosscap coefficients were obtained for $Z_2$ and
$Z_{\rm odd}$ extensions. Also
pure automorphisms due to cyclic simple currents have been considered
for boundaries \cite{fuSc5} as well as crosscaps \cite{huss2}, building 
on pioneering
work of \cite{prss2,prss3}. The general class of simple current invariants
contains, however, some additional types of invariants, such as 
automorphisms of pure extensions, and automorphisms generated by integer 
spin currents \cite{sche2}.
There are several motivations for trying to generalize the previous 
results. Simple current invariants appear abundantly in all
practical applications of RCFT to string model building, and with the
formulas we will present here, a huge set of open string models becomes
accessible to explicit computation. But in addition we expect that a
general formula will provide additional insight in the conceptual issues
involved in formulating RCFT on surfaces with boundaries and crosscaps.

Comparing the results obtained so far for pure extensions and pure 
automorphism invariants one notices a similarity between the formulas for
crosscaps. The similarity between the boundary coefficients of the two
cases is less obvious, but what they do have in common is that a crucial
r\^ole is played by the so-called
``fixed point resolution matrices". Our approach
to the problem is as follows. We start with an {\it ansatz\/} for
a general formula that includes all previous cases. This {\it ansatz\/}
consists in particular of a prescription to determine the Ishibashi
labels $m$, and the boundary labels $a$, plus a set of boundary
coefficients $B_{ma}$. We then prove that $B_{ma}$ has a left  and
a right inverse, so that it is a square matrix. This shows that the
number of boundaries equals the number of Ishibashi labels, so that the
set of boundaries is complete. We also compute the annulus coefficients
and prove that they are integral.  

Using integrality in the vacuum sector of the open string partition
function we can then, following \cite{sche11}, determine
the crosscap coefficients up to a collection of signs. Some of these signs
are fixed by imposing integrality of $K^i$; some more signs are
fixed by requiring integrality and positivity of the closed string
partition function. On the other hand, some of the signs are not 
fixed by any constraint.
They correspond to different Klein bottle choices, a possibility already
encountered in previous cases. The final check is to compute
the \Moebius coefficients and verify open string integrality.

In this letter we will only present the results of this analysis. 
Proofs and further details will be postponed to a forthcoming
publication \cite{inprep}. We begin with the description of the 
torus partition function given in \cite{krSc}. A general simple current 
invariant is characterized by a set of simple currents forming a 
finite abelian group ${\cal G}$, and a matrix $X$. The abelian
group ${\cal G}$ is a product of $k$ cyclic factors $\Zbf_{n_s}$, each
generated by some current $J_s$. The monodromy matrix $R$ of these
generators is defined as $R_{st}\,{:=}\,Q_s(J_t)$, where the monodromy
charge $Q_s$ is the combination $Q_s(i)\eq h_i\,{+}\,h_{J_s}\,{-}\,
h_{J_s i} \mod 1$ of conformal weights, plus a further constraint 
that fixes its diagonal elements modulo 2, depending
on the conformal weight of the currents.
The matrix $X$ (defined modulo 1) must satisfy 
  \be  X+X^T=R  \ee 
and a certain quantization condition on the antisymmetric part of $X$,
to be discussed below.
The matrix $X$ determines the matrix $Z\eq Z({\cal G},X)$ as follows: 
$Z_{ij}$ is equal to the number of solutions $J$ to the 
conditions\,%
 \futnote{Clearly, it is sufficient to check the 
 second condition for the cyclic group generators $J_s\iN{\cal G}$.}
  \be  \bearl  j=Ji\,,\ \ J\iN{\cal G} \ \ \hbox{~and~}\\{}\\[-.8em]
  Q_K(i)+X(K,J)=0 \mod 1  \hbox{~~for all~} K\iN {\cal G}\,.  \eear
  \labl{modinv}
Here $X(K,J)$ is the number
  \be  X(K,J) \equiv \sum_{s,t} n_s\, m_t\, X_{st}\,, \ee
with $n_s$ and $m_t$ obtained by expressing $J$ and $K$ through
the generating currents $J_s$ as $J\eq(J_1)^{n_1}\linebreak[0]
\cdots(J_k)^{n_k}$, $K\eq(J_1)^{m_1}\cdots(J_k)^{m_k}$.

The restriction to symmetric invariants implies that 
$X$ must be symmetric modulo integers. This leads to the much
simpler equation $2X\eq R$, which determines $X$ completely on the diagonal
(since $R$ is defined modulo 2), and modulo half-integers off-diagonally. 
The solutions can be described more precisely as follows.
The matrix elements $R_{st}$ and $X_{st}$ are rationals satisfying
the property that the products (no summation implied) $N_s R_{st}$, 
$N_s X_{st}$, $ R_{st} N_t$ and $ X_{st} N_t$ are integers, where
$N_s$ is the order of $J_s$. If $N_s$ is odd, $R_{ss}N_s$ is always 
even, and hence $X_{ss}$ is determined. If $N_s$ is even, $R_{ss}N_s$ 
may be odd. Then there is no solution for $X_{ss}$. In that case 
the current $J_s$ does not belong to the ``effective center", and 
cannot be used to build modular invariants. A second case in which 
$2X\eq R$ has no solutions is when $N_s$ is even and $N_s R_{st}$ 
is odd for some value of $t\,{\not=}\,s$. Then there are only 
non-symmetric invariants. In all other cases at least
one solution exists. If both $N_s$ and $N_t$ are even 
the off-diagonal element $X_{st}$ may be shifted by a half-integer. 

\mysec3{Ishibashi and boundary labels}
The modular invariant $Z({\cal G},X)$ specified by $X$
is to be multiplied with the charge conjugation matrix. Hence the
Ishibashi states correspond to the {\it diagonal\/} elements of 
$Z({\cal G},X)$, counting multiplicities. The only currents that can 
contribute are those that satisfy $Ji\eq i$. They form a group, the
stabilizer ${\cal S}_i$ of $i$. If this group is non-trivial,
multiplicities larger than 1 may occur, possibly leading to
Ishibashi label degeneracies. For pure extensions this was analysed 
in \cite{fuSc11,fuSc12}, and the conclusion is that the Ishibashi 
label degeneracy is actually equal to the fixed point degeneracy.\,%
 \futnote{This result is non-trivial because the degeneracy in the 
 extended theory is in general {\it not\/} equal to the fixed point
 degeneracy, i.e.\ the order of the stabilizer,
 but rather to the size of a subgroup, the untwisted stabilizer.} 
It is natural to extend this result to the general case, and to 
label the degeneracy by the currents that cause it. Hence our 
{\it ansatz\/} for the Ishibashi labels is
  \be  m=(i,J);   \ \  J \iN {\cal S}_i \ \hbox{~with~} \
  Q_K(i)\,{+}\,X(K,J)\eq0 \mod 1  \hbox{~~for all~}
  K\iN {\cal G} \,.  \labl{Ishi}
This {\it ansatz} produces also the correct count for pure extension 
invariants, but the labelling chosen here is not the same as in 
\cite{fuSc11,fuSc12}. 
In those papers the dual basis -- the characters $\psi_{\alpha}$
of ${\cal S}_i$ -- was used for the degeneracy labels.
This is not possible for pure automorphisms because the currents
satisfying \erf{Ishi} do not form a group in 
that case. For pure extensions, the new basis differs by a Fourier
transformation from the old one. This allows us to compute the
degeneracy metric, given the fact that it was diagonal in the
old basis. We find
  \be  g_j^{J,K}=\sum_{\alpha\beta} \psi_{\alpha}(J)\,\psi_{\beta}(K)\,
  \delta^{\alpha,\beta} = \delta^{J,K^c}\,. \ee

Now we turn to the boundary labels. The results for pure extensions
and automorphisms without fixed points is that the boundaries are in
one-to-one correspondence with the complete set of ${\cal G}$ orbits
(of arbitrary monodromy charge). As usual, fixed points lead to 
degeneracies. For pure automorphism invariants due to a half-integer
spin simple current, the degeneracy was found to be given by the order 
of the stabilizer of the orbit, whereas for pure extensions it is the 
order of the untwisted stabilizer. The latter is defined as follows 
\cite{fusS6}. For every simple current $J$ with fixed points there 
exists a ``fixed point resolution matrix" $S^J$; these matrices
can be used to express the unitary modular S-transformation matrix 
of the extended theory through quantities of the unextended theory.
The matrices $S^J$ are 
conjectured to be equal to the modular S-transformation matrices for the 
$J$-one-point conformal blocks on the torus, and are explicitly known 
for all WZW models \cite{fusS3,fusS6}, their simple current extensions 
\cite{sche10} and also for coset conformal field theories.
Elements of the matrix $S^J$ whose labels are related by the action of 
a simple current $K$ obey
  \be  S^J_{Ki,j} = F_i(K,J)\, \eE^{2\pi\ii Q_K(j)}\, S^J_{i,j} \,.  \ee
The quantity $F_i$ is called the simple current twist, and the 
untwisted stabilizer ${\cal U}_i$ is the subgroup of ${\cal S}_i$ of
currents that have twist 1 with respect to all currents in ${\cal S}_i$.
To combine the results for automorphisms and extensions, we introduce
a modified twist $F_i^X$ by
  \be  F^X_i(K,J) := \eE^{2\pi\ii X(K,J)}\, F_i(K,J)^*_{} \,,  \Labl fx
and we define the {\it central stabilizer\/} ${\cal C}_i$ as
  \be  {\cal C}_i := \{ J\iN{\cal S}_i \,|\,F^X_i(K,J)\eq1 \hbox{~for all~} 
  K \iN {\cal S}_i \} \,.  \ee
(The prescription \Erf fx is motivated as follows. The modified twist
is an alternating bihomomorphism i.e.\ obeys $F^X_i(J,J)\eq1$ for all 
$J\iN{\cal G}$. Such bihomomorphisms $F^X_i$ of an abelian group 
$\cal G$ are in one-to-one correspondence to cohomology classes 
${\cal F}^X_i$ in $H^2({\cal G},{\rm U}(1))$, thus leading to a 
cohomological interpretation \cite{bant7}. In particular, the central 
stabilizer provides a basis of the centre of the twisted group algebra 
$\Cbf_{{\cal F}^X_i}{\cal S}_i$, which also motivates its name.)

The action (by the fusion product) of the simple currents in $\cal G$
organizes the labels $i$ of the $\bar{\mathfrak A}$-theory into orbits.
Moreover, in all known cases the boundary degeneracy 
is correctly described by the order of the central stabilizer, and 
hence this is our {\it ansatz} for the general case as well. We then
choose the characters of ${\cal C}_i$ as the degeneracy 
labels. The boundaries are therefore given by
  \be  a=[i,\psi] \,,  \labl{bndlbl}
where $i$ is the label of a representative of a ${\cal G}$-orbit, and 
$\psi$ a character of ${\cal C}_i$.

\mysec4{The boundary formula}
Ishibashi states are nothing but conformal blocks for one-point 
correlation functions on the disk, i.e.\ specific two-point blocks
on the sphere. But we can think of the Ishibashi state labelled by 
$(i,J)$ also more as a {\em three\/}-point block on the sphere, 
with insertions $i,\;i^c$ and $J$. (This is actually the natural
interpretation when one wants to express such Ishibashi states in
the three-dimensional topological picture that was established in 
\cite{fffs2}.) 
Moreover, already from \cite{card9} it is known that the relation 
between Ishibashi and boundary states essentially expresses the
effect of a modular S-transformation. Together with the previous
observation, it is then natural to expect that the fixed point 
resolution matrices $S^J$ appear in the boundary coefficients.

We are therefore ready to write down the following {\it ansatz\/} 
for the boundary coefficients:
  \be  B_{(i,J),[j,\psi]} =\sqrt{|{\cal G}| \over |{\cal S}_j|\,
  |{\cal C}_j|} \, { \alpha(J)\, S^J_{i,j} \over \sqrt{S_{0,i}}}\,
  \psi(J)^*_{}\,, \labl{bnd} 
where $\alpha(J)$ is a phase to be discussed later, but which must
satisfy $\alpha(0)\eq1$. All previously studied cases are correctly 
reproduced by the remarkably simple formula \erf{bnd}. We have also 
verified that the matrix \erf{bnd} has a left- and right-inverse, given 
by $(B^{-1}_{})_{[j,\psi],(i,J)}^{}\eq S_{0,i}\,B_{(i,J),[j,\psi]}^*$. 
This establishes in particular the result that the number of boundaries 
equals the number of Ishibashi labels, i.e.\ ``completeness". This 
implies rather non-trivial relations involving the number of orbits 
of various kinds and the orders of stabilizers. 

One can also check that the annuli obtained from \erf{bnd} possess
non-negative integral expansion coefficients $A^i_{ab}$ 
with respect to the $\bar{\mathfrak A}$-characters $\chii_i$. (We 
assume, as usual, that the Verlinde formula produces non-negative
integers both for the unextended and for the extended CFT.)
When trying to express the annuli in terms of characters of 
(possibly twisted) representations of the {\em extended\/} chiral 
algebra $\mathfrak A$, one has to face the problem that their 
coefficients cannot be determined uniquely when the annuli are (as is 
usually done) considered only as functions of the variable $\tau$
associated to the Virasoro zero mode $L_0$. For reading off these
annulus coefficients unambiguously, the introduction of additional 
variables -- similar to the situation with full rather than Virasoro
specialized characters -- is required. This seems in fact to fit well 
with the above-mentioned interpretation of
Ishibashi states $(i,J)$ as three-point conformal blocks.

\mysec5{The crosscap formula}
To compute the crosscap coefficients we use the special boundary 
corresponding to the vacuum orbit, which has degeneracy 1. 
 The annulus 
coefficients for this boundary are easily found to be
  \be  A^i_{[0][0]} = \sum_{J \in {\cal G}} \delta^{Ji}_0\,. \ee
Positivity of annulus plus \Moebius strip amplitudes then requires\,%
 \futnote{The charge conjugation in the argument of
 $\eta$ is for future convenience.} 
  \be  M^i_{[0]} = \sum_{J \in {\cal G}} \eta(J^c)\, \delta^{Ji}_0\,,
  \labl{moebnul}
where $\eta(J^c)\iN\{\pm1\}$. Using the formula for the \Moebius
amplitude and the fact that the matrix $P$ is invertible, we can now
express most of the crosscap coefficients in terms of the signs $\eta$.
The result is that for fields with $Q_K(i)=0$, for all ${K \in {\cal G}}$
  \be  \Gamma_{(i,J)}={1\over | {\cal G} |} \sum_{K \in {\cal G}} \eta(K)\,
  \frac{P_{K,i}}{\sqrt{S_{0,i}}}\, \delta^{J,0} \,.  \labl{crosscap}
Note that we only get information about the $J\eq 0$ components of 
degenerate Ishibashi states,\,%
 \futnote{Note that all Ishibashi states with $J\eq0$ 
 satisfy $Q_K(i)\eq 0$  for all $K\iN{\cal G}$, so that \erf{crosscap}
 determines {\it all} such crosscap coefficients.}
because the boundary $[0]$ is itself non-degenerate. 
In \erf{crosscap} we have postulated that $\Gamma_{(i,J)}\eq0$ for 
$J\,{\not=}\,0$. This
postulate is based on known cases (where it can often be derived) and
is justified by the consistency of the resulting Klein bottle. Comparison 
of the formula for the \Moebius strip amplitude with 
\erf{moebnul} yields more information than just \erf{crosscap}. We also find
that the right-hand side of \erf{crosscap} must vanish if $Q_K(i)\not=0$  
for some ${K \in {\cal G}}$. This implies relations between the
signs $\eta(J)$. They can be derived using the relation
  \be  \bearl
  P_{i,K^{2\ell} j}=\rho(\ell)\, \eE^{\pi\ii\Delta(2\ell,j)}
  \eE^{2\pi\ii\ell Q_K(i)}P_{ij} \\{}\\[-.7em]
  {\rm with}\ \ \Delta(\ell,i)
  = h_{K^{\ell}i}\,{-}\,h_{K^{\ell}}\,{-}\,h_i\,{+}\,\ell Q_K(i) \,, \ \
  \rho(\ell) = \eE^{\pi\ii (r\ell +M_{2\ell})} , \ \
  M_{\ell} = h_{K^{\ell}}\,{-}\,{r{\ell}(N-{\ell})\over 2N}  \eear
  \labl{TheFormula}
for the matrix elements of $P$ ($N$ is the order of the current $K$). 
The number $\rho(\ell)\,\eE^{\pi\ii\Delta(2\ell,j)}$ is a sign, and 
the factors
$\eta(J)$ must be chosen such that they cancel these signs. This is
necessary and sufficient to ensure the vanishing of the right-hand side
of \erf{crosscap} for some of the charges: 
namely all charges with respect to currents
$K$ that can be written as a square, $K\eq L^2$ for some $L\iN{\cal G}$. 
These currents form a subgroup ${\cal G}_{\rm E}$
of ${\cal G}$, and will henceforth be called {\it even}
currents. Note that any current of odd order is even. Vanishing
of the expression for the 
remaining charges then turns out to yield no further conditions. This
follows from the fact that $P_{ij}\eq0$ if $i$ and $j$ have different 
charges with respect to a (half)-integer spin current of order 2. 
Since there is no further condition, the signs $\eta(J)$ remain
unconstrained on the cosets ${\cal G}/{\cal G}_{\rm E}$.

The precise relation that the coefficients $\eta$ have to satisfy 
can be written more conveniently by defining
  \be  \beta(J) := \eE^{\pi\ii h_J} \eta(J) \,.  \labl{etabeta}
We then find that for even currents $K=L^2$ (and any current $J$)
\futnote{The numbers $\eE^{-2\pi\ii Q_K(J)}$ furnish a
two-cocycle on the quotient group ${\cal G}/{\cal G}_{\rm E}$.
Formula \erf{betabetaeven} thus means that $\beta$
forms a one-dimensional representation of the corresponding twisted
group algebra, which is possible only when the cocycle is a
coboundary; this is indeed the case.}
  \be  \beta(KJ)=\beta(K)\beta(J)\, \eE^{-2\pi\ii Q_K(J)}=
  \beta(K)\beta(J)\, \eE^{-2\pi\ii X(K,J)}\,. \labl{betabetaeven}

\mysec6{Integrality and positivity}
We can now compute the Klein bottle and check integrality 
and positivity in the closed sector. It turns out that there are 
no further constraints as long as
there are no fixed points. If, however, we assume that all allowed types
of orbits actually do occur, in order to obtain a formula that is valid
in all cases, then a further constraint is necessary, namely 
  \be \beta(KJ)=
  \beta(K)\beta(J)\, \eE^{-2\pi\ii X(K,J)}\,; \labl{betabetaodd}
this is identical to \erf{betabetaeven}, but this time also valid
for odd currents. The number of free signs is therefore as follows.\,%
 \futnote{The overall sign of the crosscap coefficients is
 always free, and can be fixed by choosing $\eta(0)\eq1$.}
The number of cosets ${\cal G}/{\cal G}_{\rm E}$ is $2^M$, with $M$ the
number of even cyclic factors of ${\cal G}$.
Within each coset the signs $\eta(J)$
can be related using \erf{betabetaeven}, but signs in different 
cosets are unrelated, so that there is a total of $2^M$ sign choices. 
If \erf{betabetaodd} is valid, all signs can be expressed in terms of those
of the generators of the even cyclic factors. This reduces the number
of sign choices to $M$. Since this is the generic solution, it is
the one most likely to survive further consistency checks, but we cannot
rule out the possibility that more general sign choices are permitted
in theories where certain a priori allowed fixed
points simply do not occur. 

The last consistency condition follows from positivity and integrality
of the open string sector, and concerns the phases $\alpha(J)$
introduced in \erf{bnd}. These phases do not appear in the \Moebius 
strip partition function, and enter the annulus only as the combination
$\alpha(J)\alpha(J^c)$. Such phases already occurred for the $\Zbf_2$
extensions and automorphism invariants discussed in \cite{huss2} and 
\cite{husc}, where they were found to be related to the sign choices 
in the crosscap. The same is true here, the precise relation being
  \be  \alpha(J)\,\alpha(J^c)= \beta(J) \,.  \labl{alphaalphabeta}
If $J$ has fixed points it either has integer or half-integer spin.
Since $\eta(J)$ is a sign, it follows from \erf{etabeta} that 
$\beta(J)$ is a sign for integer spin currents, and $\pm\ii$ for 
half-integer spin currents. If we fix the convention 
$\alpha(J)\eq\alpha(J^c)$,\,%
 \futnote{For WZW models $S^J\eq S^{J^c}$, which makes it 
 natural to impose the same condition on the phases.} 
we find that $\alpha(J)$ is a fourth root of unity for integral spin 
currents, and a primitive eighth root of unity for half-integer spin 
currents. This resolves another apparent conflict between the earlier 
results for pure extensions and automorphisms. Namely, in the formulas of 
\cite{fuSc11} for the former case the matrices $S^J$ appear, whereas 
in the automorphism case in \cite{fuSc5} a slightly different matrix 
appears, namely the modular transformation matrix of the relevant
``orbit Lie algebra" that was defined in \cite{fusS3}. Its definition 
involves folding a Dynkin diagram, a procedure that is  
only available for WZW models. In that case, the matrix differs from
the fixed point resolution matrix $S^J$ by a primitive eighth root of 
unity, if $J$ has half-integer spin, and by a fourth root of unity if 
$J$ has integer spin. The present formalism allows us to use
$S^J$ in all cases; it has the additional advantage that $S^J$ has
a more general definition, and has been computed in more cases.

\mysec7{More general solutions}
There is (at least) one further generalization possible whenever 
the RCFT under consideration has an additional simple current $K$ 
that is not contained in ${\cal G}$. We can then generalize the 
results of \cite{huss} to obtain 
different Klein bottle projections and correspondingly different
boundary coefficients. It turns out that $K$ must satisfy the constraint
  \be  Q_J(K)=0 \hbox{~for all~} J\iN{\cal G} \hbox{~with~}
  J^2\eq 0 \,. \ee
The modified formula for the boundaries is
  \be  B_{(i,J),[j,\psi]} = 
  \sqrt {|{\cal G}| \over |{\cal S}_j|\,|{\cal C}_j|}\, {\alpha(J)\,
  S_{i,j}^J \over \sqrt{S_{K,i}}}\, \psi(J)^*_{}\,, \ee
and for the crosscap coefficients we find\,%
 \futnote{As explained in \cite{huss}, any ambiguity in the 
 choice of the square roots cancels out in the amplitudes.}
  \be  \Gamma_{i,J}={1\over \sqrt{|{\cal G}|}}\, {1\over \sqrt{S_{K,i}}}
  \sum_{L \in {\cal G}} \eta(K,L)\, P_{KL,i}\,\delta^J_0 \,.  \ee
The effect of the ``Klein bottle current" $K$
is to flip some signs of the Klein bottle projection. One finds nothing new
(up to a permutation of the boundaries) if $K$ is the square of another
current, or if $K\iN{\cal G}$. The signs $\eta(K,L)$ are given by
  \be  \eta(K,L)=\eE^{\pi\ii (h_K -h_{KL})}\beta(L)\,. \labl{etabetaK}
The coefficients $\beta(L)$ must satisfy the same  condition 
\erf{betabetaodd} as in the case $K\eq 0$, 
and the coefficients $\alpha(J)$ are related
to phases $\beta(L)$ as in \erf{alphaalphabeta}. Note that although
the coefficients $\beta$ satisfy the same product formula independently
of the choice of the Klein bottle current $K$, the solutions {\it do}
depend on $K$ because of the additional requirement that $\eta(K,L)$
must be $\pm1$. The phases $\alpha(L)$ are relevant only if $L$ fixes
some field. Then $h_L$ is integer or half-integer and 
$Q_K(L)=h_K+k_L-h_{KL} \mod 1 = 0 \mod 1$. From \erf{etabetaK}
we then find that $\beta(L)=\eE^{\pi\ii h_L} \eta(K,L)$. Hence for any choice
of $K$ the coefficients $\alpha(J)$ are fourth (eighth) roots of unity of
integer (half-integer) spins, as before. 

\mysec8{Summary}
The main results of this paper are formulas \erf{Ishi} and
\erf{bndlbl}, which specify the Ishibashi and boundary labels,
as well as \erf{bnd} and \erf{crosscap}, which provide the boundary 
and crosscap coefficients, for a general simple current modular 
invariant that is based on the charge conjugation invariant. 
(In addition, the phases appearing in these expressions are subject to 
the constraints \erf{betabetaodd} and \erf{alphaalphabeta}.)
We do not have a proof that these results lead to consistent 
correlation functions on arbitrary Riemann surfaces. However, 
they do satisfy a set of quite non-trivial consistency conditions 
at the one-loop level, as well as the completeness conditions.
Their simplicity and generality strongly suggest that this must
indeed be the correct answer.

\small
 \newcommand\wb{\,\linebreak[0]} \def\wB {$\,$\wb}
 \newcommand\Bi[1]    {\bibitem{#1}}
 \newcommand\J[5]   {{\sl #5}, {#1} {#2} ({#3}) {#4} }
 \newcommand\Prep[2]  {{\sl #2}, preprint {#1}}
 \def\jf    {J.\ Fuchs}
 \def\adma  {Adv.\wb Math.}
 \def\anop  {Ann.\wb Phys.}
 \def\aspm  {Adv.\wb Stu\-dies\wB in\wB Pure\wB Math.}
 \def\atmp  {Adv.\wb Theor.\wb Math.\wb Phys.}
 \def\comp  {Com\-mun.\wb Math.\wb Phys.}
 \def\ijmp  {Int.\wb J.\wb Mod.\wb Phys.\ A}
 \def\jhep  {J.\wb High\wB Energy\wB Phys.}
 \def\mpla  {Mod.\wb Phys.\wb Lett.\ A}
 \def\nuci  {Nuovo\wB Cim.}
 \def\nupb  {Nucl.\wb Phys.\ B}
 \def\phlb  {Phys.\wb Lett.\ B}
 \def\phrl  {Phys.\wb Rev.\wb Lett.}
 \def\NH     {{North Holland Publishing Company}}
 \def\SV     {{Sprin\-ger Ver\-lag}}
 \def\WS     {{World Scientific}}
 \def\Ad     {{Amsterdam}}
 \def\Be     {{Berlin}}
 \def\Si     {{Singapore}}

\end{document}